# Comparative study between three models for the levitation height calculation

Nabin K. Raut[1,*], Jeffery Miller[1], Raymond Y. Chiao[1], Jay E. Sharping[1]

1 University of California, Merced

* Corresponding Author: nraut@ucmerced.edu

**Abstract.** Magnetic levitation has been demonstrated and characterized within the coaxial microwave cavity [1,2]. A permanent neodymium magnet is levitated from the edge of the finite size superconductor [3,4]. One challenge is developed a better method to calculate levitation height [5]. This paper compares three models, Mirror method, finite-size superconductor, and two-loop model, for the levitation height calculation. The limitations and advantages of each model is discussed in detail.

## INTRODUCTION

An object levitates when the upward pushing levitation force balance out the downward pulling gravitational force. The levitated system is a freely floating system free from the clamping and surface contact. This degree of freedom makes is it free of losses induce due to the friction [6]. Few examples of levitated systems are optical levitation of the dielectric particle, levitating train, and superconducting magnetic levitation [7,8].

Superconducting cavities already proven to achieve the high-gradient and excellent quality factor performance goals [9]. Importantly, ultra-low damping of the Meissner-levitated ferromagnet within the superconducting lead trap (Type-I superconductor) is demonstrated [10]. For a magnet to levitate above a superconductor, the upward pulling force is provided by the Meissner force [11]. The Meissner effect is associated with the complete expulsion of the magnetic flux from the interior of the superconductor [12]. Here necessary condition being the applied field is less than the critical field of the superconductor. The boundary condition is that the perpendicular component of the magnetic field will be zero on the superconductor's surface [13]. Based on this boundary condition, different models have been developed to calculate levitation force (or energy). One such model is the mirror method. The mirror method considers the magnet and its diamagnetic image as a dipole [14]. However, it does not consider the dimension of either the superconductor or the magnet. A model that is developed by Lugo et al. [15] considered the size of the superconductor. They considered the magnet a point dipole and the superconductor a continuous array of point dipoles. The levitation force (or energy) is then obtained by integrating the dipole-dipole interaction between the real and the image magnet over the volume of the superconductor [15,16]. The main limitation in their calculation is exclusion of the magnet's size.

One challenge of the Meissner levitation is the edge effect, where the magnet is pushed toward the edge of the superconductor. In this this type of levitation the center of the superconductor does not coincide with the axis of the magnet. It is called non-coaxial case. The two methods discussed above are limited to the co-axial case only. In this study, a model is developed using two-loop model to calculate the levitation height [17]. This model considers the size of both the magnet and the superconductor and also valid for the non-coaxial case. We find that the two-loop model calculates levitation height more accurately than the mirror method and the method developed by Lugo et at.

## MIRROR METHOD

The schematic of the assumption of the mirror method is shown in Fig. 1. For the type I superconductor, this model assumes that the magnet on the superconductor has its diamagnetic image inside the superconductor. Both magnets (real and image) move in the opposite direction. The interaction between the magnets is most robust near the superconductor's surface and becomes weaker as it goes farther from the surface.

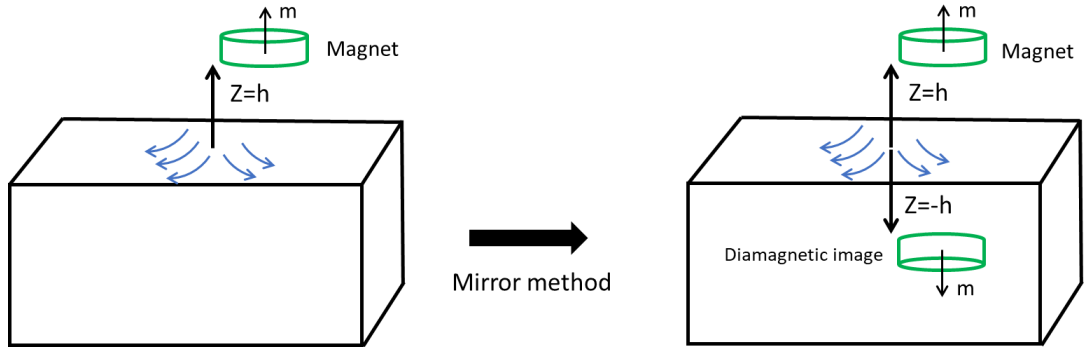

**FIGURE 1.** The mirror method's schematic view of a magnet at a height h above a superconductor.

The mirror method can calculate the potential energy (or levitation force) calculation. Consider a permanent magnet and its corresponding image as an induced dipole. The potential energy on the magnet due to the induced dipole depends upon the magnetic moment ($\vec{m}$) and magnetic field ($\vec{B}$). Mathematically the potential energy can be written as:

$$\vec{U} = \frac{1}{2}(\vec{m} \cdot \vec{B}) \qquad (1)$$

We can calculate a magnetic field due to the dipole at the distance z from its center is given by the relation below [18]:

$$\vec{B}(0, 0, z) = \frac{\mu_0 m}{4\pi} \frac{2}{z^3} \hat{k} \qquad (2)$$

Substituting the above expression into Eq. (1) results in the potential energy due to the force exerted by the mirror image on the permanent magnet.

$$U(0,0,z) = \frac{\mu_0 m^2}{4\pi} \frac{(1 + sin^2\theta)}{(2z)^3} \qquad (3)$$

Here, $(1 + sin^2(\theta))$ takes account of the angle of the magnet with the superconductor and $(2z)$ is the distance between the real and image magnet. The potential energy due to the radially magnetized magnet is half that of the axially polarized magnet. The total potential energy includes the gravitational $(Mgz)$ potential energy as well.

$$U_{total}(0,0,z) = \frac{\mu_0 m^2}{4\pi} \frac{(1 + sin^2\theta)}{(2z)^3} + Mgz \qquad (4)$$

The above expression for the potential energy assumes the magnetic field is completely expelled from the superconductor, an infinite plane. The magnet levitates at the point with the least potential above the superconductor [19].

Eq. (4) is plotted for an axially magnetized N52 permanent neodymium magnet of a radius and height of 0.5 mm in Fig. 2. Here $\theta = 90°$ is used. The total potential energy near the superconductor is high because of the large repulsion between the magnet and its diamagnet image. As the magnet moves farther away from the superconductor's surface, the potential energy quickly falls off. Its value becomes minimum at 3.8 mm above the superconductor. Hence, the magnet levitates at this minimum energy point.

The levitation force than can be calculated as [19]:

$$F_{Lev}(0,0,z) = -\Delta U_{Lev}$$
$$F_{Lev}(0,0,z) = \frac{6\mu_0 m^2}{4\pi} \frac{(1 + sin^2\theta)}{(2z)^4} \qquad (5)$$

The vertical stiffness can be derived from equation (5) as:

$$K_z = -\frac{\partial F_{Lev}(0,0,z)}{\partial Z}$$
$$= \frac{48\mu_0 m^2}{4\pi} \frac{(1 + sin^2\theta)}{(2z)^5} \qquad (6)$$

This will lead to the resonance frequency of:

$$\omega_z = \sqrt{\frac{K_z}{M}}$$

$$\omega_z = m\sqrt{\frac{3\mu_0(1 + sin^2\theta)}{8\pi * M} \frac{1}{z_0^5}}$$

$$\omega_z = \sqrt{\frac{4g}{z_0}} \qquad (7)$$

For, an N52 neodymium magnet of mass (M) 2.75 *milligram* and magnetic moment (m) 0.46 *(mA)m²* levitating at a distance 3.8 *mm* above the superconductor, $f_z$ will be ≈11 *Hz*.

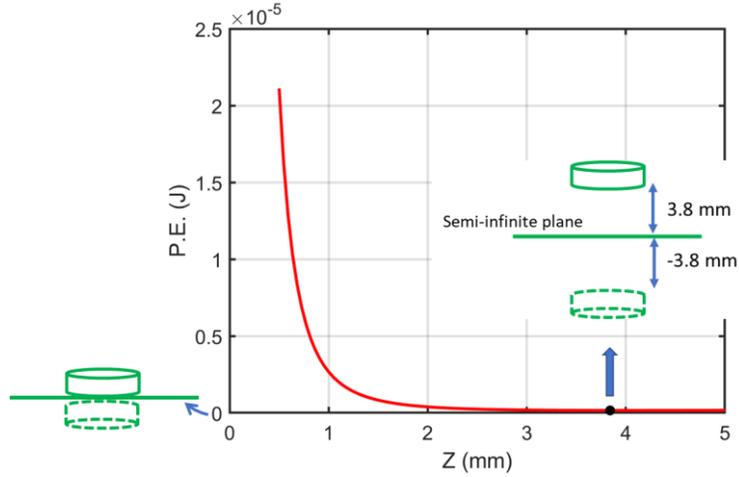

**FIGURE 2.** Total potential energy as a function of the vertical position of the magnet. The calculation starts from 0.5 mm vertical height. The substantial repulsion between the real and image magnets results in enormous energy in this distance.

## FINITE-SIZE SUPERCONDCUTOR

The main drawback of the Mirror method is that it considers the superconductor as an infinite size. The approach that Lugo et al. [15] took considered the size of the superconductor in their calculation. The levitation force due to a superconducting cylindrical of radius R and thickness t, can be written as:

$$F_{Lev} = \frac{\mu_0}{4\pi} \frac{3m^2(1 + \sin^2\theta)}{32} [f(a) - f(a+t)] \tag{8}$$

Where:

$$f(z) = \frac{1}{z^4} - \frac{5R^2 + 3z^2}{3(R^2 + z^2)^3} \tag{9}$$

In Fig. 3, levitation force is calculated as a function of the vertical position of the magnet of the same dimension and strength that have been used for Fig. 2 generation. For the size of the superconductor, the dimension of the stub, $R = 2\ mm,$ and $t = 5\ mm$, is used. The levitation force balances the gravitational force at 2.75 mm above the superconductor. This value is 28 percent smaller than the levitation height predicted by the mirror method.

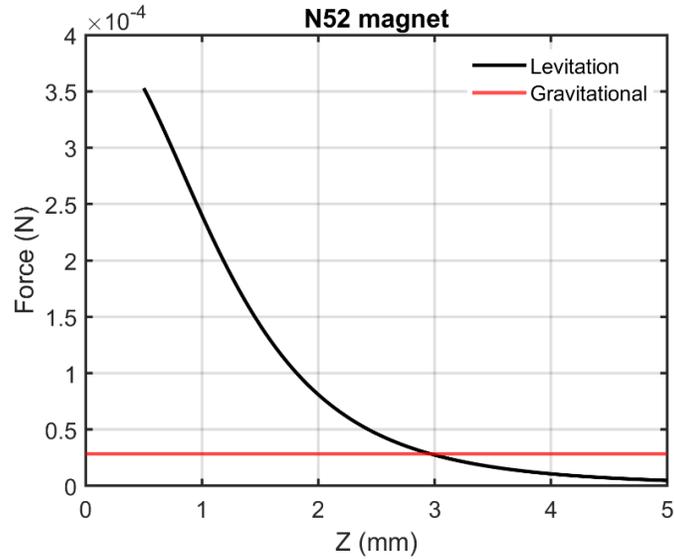

**FIGURE 3.** Levitation force for a finite-size superconductor. Here, 0.25 mm (half the thickness of the magnet) is subtracted from 3 mm to get levitation height from the surface of the superconductor to the surface of the magnet. The upward levitation force balances the downward gravitational force at 2.75 mm.

## Two-loop model

In the levitation experiment, the axis of the magnet might not necessarily coincide with the axis of the superconductor. Similarly, the size of the magnet is comparable to the size of the superconductor. Therefore, the methods discussed above result in inaccurate levitation height calculations [20].

We have used a two-loop model to calculate levitation height for the magnetic levitation from the stub of the cavity. As shown in Fig. 4, the magnet and its image are considered two current loops in this model. Their distance is taken from the center of mass of the magnet. Let us suppose the magnet has a radius and height of $R_M$ and $h$, respectively. Similarly, the superconducting stub has a radius of $R_S$ and thickness of $t$. In the two-loop model, the magnet is replaced by a loop of current with the same radius as a magnet. Also, a loop of current replaces the image magnet with a radius equal to the radius of the superconductor. The distance between the two loops is now $2Z + h$ instead of $2Z$ between two magnets.

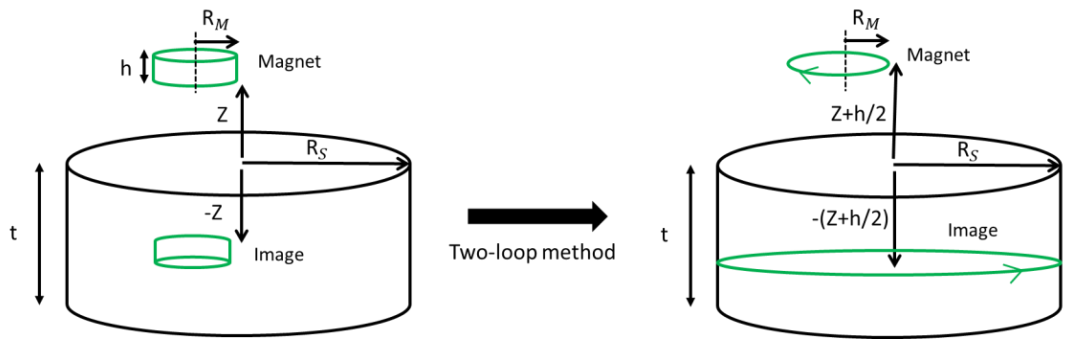

**FIGURE 4.** Two-loop representation of the magnet and its image. Here, the magnet and its image are replaced by current-carrying loops in the opposite direction.

Now, let's calculate the magnetic field due to the two loops. The vector potential has only an azimuthal component, which is given by the equation:

$$A_\emptyset = \frac{\mu_0}{4\pi}[(R_S+r)^2+z^2]^{\frac{1}{2}} \cdot \left[\left(1-\frac{1}{2}k^2\right) \cdot K(k) - E(k)\right] \tag{10}$$

Where:

$$k(r) = \frac{4R_S r}{(R_S+r)^2+z^2} \tag{11}$$

$$r = [(R_M\cos\emptyset_2+y)^2+(R_M\sin\emptyset_2)^2]^{\frac{1}{2}} \tag{12}$$

$$R = [R_S^2+r^2+z^2-2R_S r\cos\emptyset_1]^{\frac{1}{2}} \tag{13}$$

Using, relation $B = \Delta \times A$, we get:

$$B_z = \frac{\mu_0 I}{4\pi[(R_S+r)^2+z^2]^{\frac{1}{2}}}\left[\frac{R_S^2-r^2-z^2}{(R_S-r)^2+z^2} \cdot E(k) + K(k)\right] \tag{14}$$

$$B_r = \frac{\mu_0 I z}{4\pi r[(R_S+r)^2+z^2]^{\frac{1}{2}}}\left[\frac{R_S^2+r^2+z^2}{(R_S-r)^2+z^2} \cdot E(k) - K(k)\right] \tag{15}$$

$$B_\emptyset = 0 \tag{16}$$

$$m = \pi I R_S^2 \tag{17}$$

From the frame of reference of the magnet, the components of the magnetic field will be:

$$B_{r|x'-y'-z'} = B_{r|x-y-z}\cos\left(\tan^{-1}\left(\frac{R_M\sin\emptyset_2}{y+R_M\cos\emptyset_2}\right)-\emptyset_2\right) \tag{18}$$

$$B_{\emptyset|x'-y'-z'} = B_{\emptyset|x-y-z}\sin\left(\tan^{-1}\left(\frac{R_M\sin\emptyset_2}{y+R_M\cos\emptyset_2}\right)-\emptyset_2\right) \tag{19}$$

Now potential energy will be the dot product of the magnetic moment of the magnet ($\vec{m}$) and response field from the image magnet $\vec{U} = \frac{1}{2}(\vec{m} \cdot \vec{B_Z})$. Using equation (1) the potential energy yields:

$$U = \frac{\mu_0 I m}{4\pi[(R_S+r)^2+z^2]^{\frac{1}{2}}}\left[\frac{R_S^2-r^2-z^2}{(R_S-r)^2+z^2} \cdot E(k) + K(k)\right] \tag{20}$$

The potential energy calculated from the two-loop model is compared with the potential energy calculated by using the mirror method and finite-size superconductor in Fig. 5. There is a large discrepancy of energy between three models on the axis of the superconductor. The main reason for such a large discrepancy is that the two-loop model more accurately takes account of the size of the superconductor and the magnet.

Furthermore, in comparison, the two-loop model calculates levitation height more accurately than the other two models. As shown in Fig. 5 (see three markers), the two-loop model calculates levitation height 2.65 *mm*. It is smaller than the value predicts by the mirror method (3.8 *mm*) and the finite SC method (2.75 *mm*). The mirror method and finite-size superconductor overestimate the levitation height, respectively, by 31 and 4 percent.

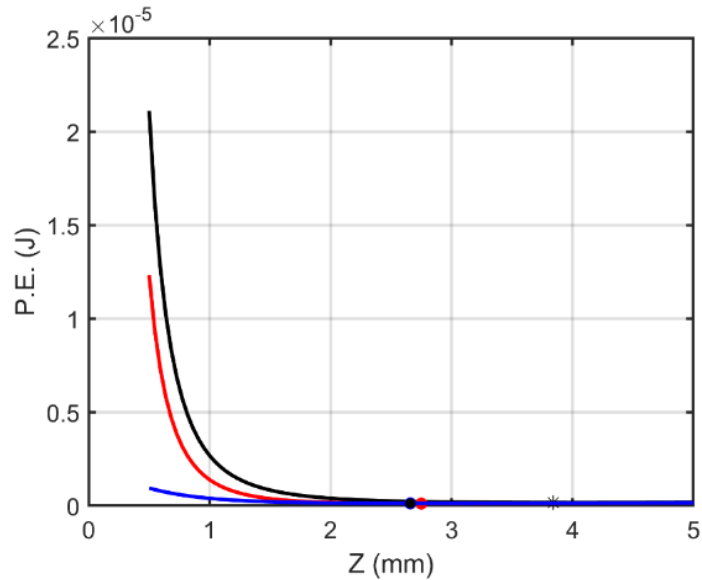

**FIGURE 5.** Comparison of the potential energy calculated at the center of the superconductor using three models: the mirror, finite-size superconductor, and two-loop model.

Figure 6 compares levitation height between the three models as a function of the strength of the magnet. In all three models, a common trend is levitation height increases with the strength of the magnet. The stronger magnet produces the stronger image magnet. Hence, the larger levitation force results in the greater levitation height.

The significant difference in levitation height calculations between the three models comes from the weaker magnets ($B_r < 1\,T$) at the center of the superconductor. The difference between the finite SC and two-loop model gets smaller as remanence goes higher than 1 T, reaching zero above 1.5 T.

The significance difference between three models appears at the edge of the superconductor. The Mirror and finite SC model considers only the coaxial superconductor-magnet case, where the magnet's axis coincides with the axis of the superconductor. However, levitation height is reduced as the magnet moves from the center to the edge of the superconductor. For example, according to the two-loop model, for the magnet with a strength of 1.5 T, the levitation height is reduced from

2.7 mm to 2 mm as we go from the center to the edge of the superconductor. The main reason is the reduction of the response supercurrent at the edge of the superconductor than that at its center. The two-loop model is the only model among three models to calculate accurate levitation height in this case.

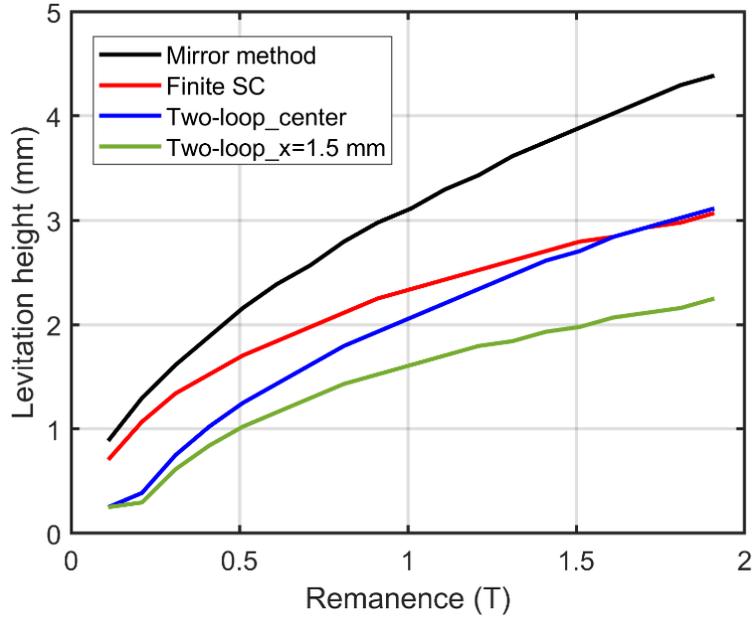

**FIGURE 6.** Levitation height as a function of remanence of the magnet. Three models (mirror, finite SC, and two-loop) are compared at the center and edge of the superconductor.

## CONCLUSION

Comparative study between three models, the mirror method, finite-size superconductor, and two-loop model, has been done. For the coaxial case, discrepancy between three models is less for the case of weaker magnets. Such discrepancy increases drastically for the stronger magnet. In the experiment, a magnet could levitation from the edge of the superconductor [21]. In that case, the two models vary drastically from the two-loop model. In conclusion, the two-loop model is the better method for this type of magnetic levitation.

## ACKNOWLEDGEMENT

I would like to thank Jefferson National Lab for giving me preparation time for this manuscript. Appreciation goes to Prof. Chih-Chun Chien, Prof. Michael Scheibner, Dr. Pashupati Dhakal for all help and support.